\documentstyle[11pt,paspconf,epsf]{article}

\begin{document}

\title{Properties of the Neutral ISM in NGC 185 and NGC 205}
\author{L. M. Young}
\affil{Astronomy Dept., New Mexico State University, P.O. Box 30001,
Las Cruces, NM 88003, USA}

\begin{abstract}
We describe the properties and kinematics of the neutral gas in
the two dwarf elliptical galaxies NGC 185 and NGC 205, companions of M31.
The kinematics of the HI suggest that an internal 
origin of the gas (i.e. stellar mass loss) should be considered, at least in the
case of NGC 185.
The molecular clouds in these dwarf ellipticals are similar to 
Galactic giant molecular clouds in many ways.
However, the HI column density in the dwarf ellipticals is surprisingly
low.
We suggest that a lower interstellar UV field
may make it easier to form and retain molecular
gas in the dwarf ellipticals than in the solar neighborhood.
\end{abstract}


\section{Introduction}

NGC~185 and NGC~205 are unique in being among the
the closest early-type galaxies ($<$ 1 Mpc) that have a neutral
interstellar medium (ISM).
They are small galaxies (luminosities of a few $\times 10^8$ L$_\odot$),
and they are called dwarf ellipticals or dwarf spheroidals
in order to distinguish them from ``regular" or ``giant" ellipticals which follow a
different luminosity-size relation (Ferguson \& Binggeli 1994).
Their proximity means that we can image their ISM at
resolutions of $\sim$50~pc and better, on the scales of individual 
molecular clouds.
Individual stars can be resolved, and color-magnitude diagrams provide
information on the star formation history of these galaxies and their
inventory of young massive stars (e.g. Martinzez-Delgado in this volume;
Peletier 1993; Lee et al. 1993; Lee 1996).
Thus, in NGC~185 and NGC~205 we can probe the star-gas ecosystem in 
detail that simply isn't possible with any other early-type galaxy.
Such observations give valuable perspective on how the ISM works in an
environment unlike the spirals we are familiar with; they also give clues
to the origin and evolution of the gas in early-type galaxies.

\section{Neutral Gas Content}

\begin{figure}
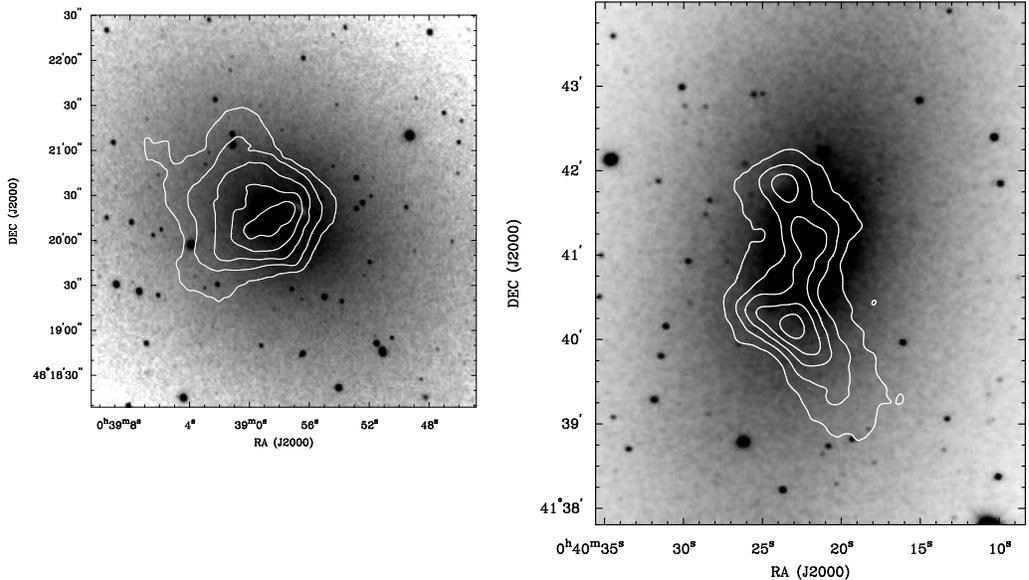

\plotfiddle{185stars.ps}{4 in}{-90}{30}{30}{-230}{300}
\plotfiddle{205stars.ps}{4 in}{-90}{40}{40}{-70}{610}
\vspace{-5.4 in}
\caption{Distribution of the HI in NGC~185 (left) and NGC~205 (right).
Greyscales are optical images of the galaxies;
contours in white are the HI column densities at (10, 30, 50, 70, 90)
percent of the peak.  Peak HI column densities are 
$3\times 10^{20}$~cm$^{-2}$ for NGC~185 and 
$4\times 10^{20}$~cm$^{-2}$ for
NGC~205.  From VLA observations by Young \& Lo (1997).
\label{stars+hi}
}
\end{figure}

For many years it has been known that NGC~185 and NGC~205 contain dust,
which is visible in optical images (Hodge 1963, 1973).
Recent radio observations show that they also contain a few 
$\times 10^5$ M$_\odot$ of atomic and molecular gas (Table~\ref{table1}).
Figure~\ref{stars+hi} shows the distributions of HI
in these galaxies. 
All of the neutral ISM is centrally concentrated and is found
within 450 pc (for NGC~205) or 180 pc (for NGC~185) of the centers of the
galaxies.
In this respect, the ISM of the dwarf ellipticals is significantly 
different from the
neutral gas in ellipticals, a topic which will be discussed further
in the next section.

\begin{table}
\caption{Masses of ISM components in NGC~185 and NGC~205}\label{table1}
\begin{center}
\begin{tabular}{lccl}
 & NGC 185 & NGC 205 & Notes \\
\tableline
L$_B$ (L$_\odot$) & $1.1\times 10^8$  & $3.4\times 10^8$ & 1 \\
M(HI) (M$_\odot$) & $1\times 10^5$  & $4\times 10^5$ & 2 \\
M(H$_2$) (M$_\odot$) & $4\times 10^5$  & $1-2\times 10^5$ & 3,4  \\
M(dust) (M$_\odot$) & $7\times 10^2$ & $2-4\times 10^3$ & 5,6  \\
\tableline
\end{tabular}
\end{center}
Notes: (1) RC3, assuming distances of 0.6 Mpc for NGC~185 and 0.85 Mpc
for NGC~205. 
(2) Young \& Lo (1997).
(3) Welch, Mitchell, \& Yi (1996); uses a CO--H$_2$ conversion factor that is
four times higher than that for NGC~205.
(4) Welch, Sage, \& Mitchell (1998).
(5) Roberts et al. (1991), rescaled to a distance of 0.6 Mpc.
(6) Fich \& Hodge (1991), rescaled to a distance of 0.85 Mpc; lower limit
(doesn't cover the entire dust distribution).
\end{table}

The mass of the interstellar medium in NGC~185 and NGC~205 is, in the
following sense, typical.
NGC~185 and NGC~205 have HI mass-to-blue luminosity ratios of about
$10^{-3}$.
Both Knapp et al. (1985) and Huchtmeier et al. (1995) find that substantial numbers,
perhaps half, of all ellipticals contain HI with an M(HI)/L$_B$ ratio of
$10^{-3}$ or higher.
Of course, the relationship between the
interstellar media of dwarf ellipticals and giant ellipticals
is poorly understood because
the number of dwarf ellipticals with detected HI is so small, and because 
dwarf and giant ellipticals may have had different evolutionary
histories.

\section{Kinematics and some clues about the origin of the ISM}
\label{kinematics}

HI synthesis observations of elliptical galaxies provide vital 
clues to the origin of the neutral gas in those galaxies. 
The HI in ellipticals is typically highly extended with respect to the optical
galaxies and is rotating (van Gorkom 1992 and references therein; see also
Morganti in this volume).
The stars in these galaxies have much lower angular momenta per unit mass
than the gas does.
These facts constitute the most important pieces of evidence that the
neutral gas in elliptical galaxies probably has an external origin (was
captured in an interaction), as opposed to an internal origin (stellar mass
loss).

Velocity fields of the HI in NGC~185 and NGC~205 are presented by Young \& Lo 
(1997).
In the case of NGC~205, the elongated gas distribution appears to be
rotating at 20 km~s$^{-1}$ about its minor axis.
The stars are not rotating, with an upper limit of 2 km~s$^{-1}$ from
observations out to 150\arcsec\ along the major axis at $-$15$^\circ$ 
(Bender et al. 1991).
As in the case of giant ellipticals, the kinematics of the HI
are inconsistent with the kinematics of the stars.
The gas in NGC 205 may also have had an external origin;
on the other hand, Sage et al. (1998) have argued that
it may have had an internal origin and been torqued by some interaction
with M31.


The velocity field of NGC~185 shows no sign of rotation.  In fact, the
HI in NGC~185 is distributed in a handful of distinct clumps which appear
to be moving with disordered velocities with a dispersion of 14 km~s$^{-1}$
(Young \& Lo 1997).
The stars also do not appear to be rotating; they have a velocity
dispersion of 22 km~s$^{-1}$\ (Bender et al. 1991; Held et al. 1992).
Thus the kinematics of the HI in NGC~185 are {\it consistent}
with the kinematics of the stars, suggesting that (barring some perverse
inclination effects) an {\it internal origin}
should be considered for the gas in NGC~185.

It has long been expected that stellar mass loss should contribute
some significant amount of gas to the ISM of elliptical galaxies (e.g. Faber \&
Gallagher 1976).
Using Faber \& Gallagher's mass loss rate, it would have taken only
$10^9$ years to accumulate the ISM in NGC~185 from stellar mass loss. 
We speculate that the neutral ISM in these dwarf ellipticals is related to
the hot (X-ray emitting) ISM in giant ellipticals in this way:
the stellar mass loss in a giant elliptical can be heated to X-ray
temperatures by the velocity dispersion of the stars.
The stellar velocity dispersions are much lower in dwarf ellipticals, and
perhaps the stellar mass loss could remain neutral.

\section{Properties of the ISM}

\begin{figure}
\plotfiddle{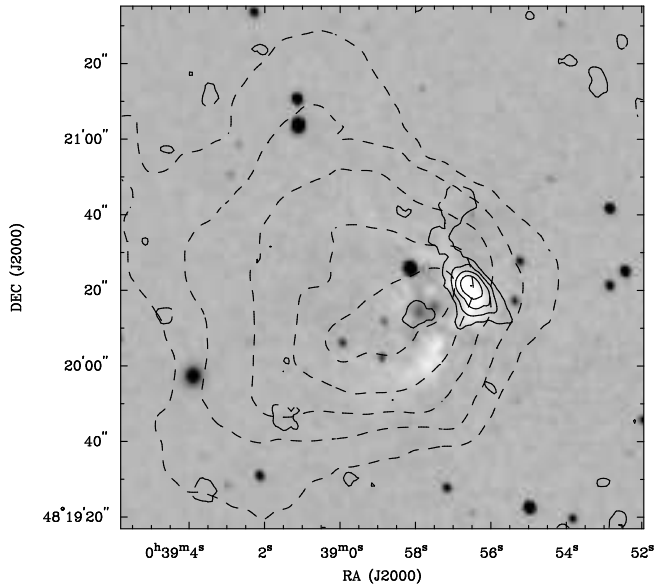}{3 in}{-90}{40}{40}{-150}{270}
\vspace{-0.6 in}
\caption{
Distributions of HI, CO, and dust in NGC 185.  The greyscale is a
B image of NGC~185 (courtesy of L. van Zee), which has had a smooth
elliptical model subtracted.  Patches of dust are white.
The dashed contours show the integrated HI intensity as in
Figure~\ref{stars+hi}, and the 
solid contours show integrated CO intensity obtained with the BIMA array
(Young 1998).
No CO emission is detected from the dust cloud southeast of the galaxy
center, but that may be due to limited sensitivity (see also Welch et al.
1996 and Young \& Lo 1997).
\label{185dust}
}
\end{figure}

\begin{figure}
\plotfiddle{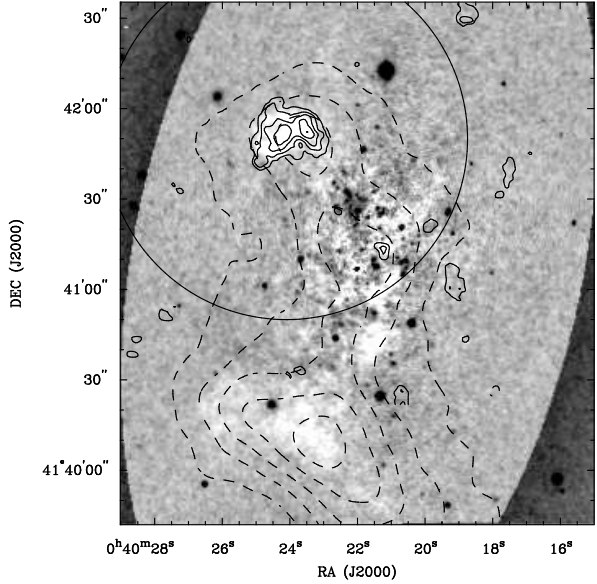}{3 in}{-90}{40}{40}{-170}{270}
\vspace{-0.6 in}
\caption{HI, CO, and dust in NGC 205, similar to Figure~\ref{185dust}.
The CO image is from BIMA observations of Young \& Lo (1996).
R. Peletier kindly provided a B image of NGC~205 (Peletier 1993).
The BIMA field of view, shown by the large circle, does not cover all of
the CO emission in the galaxy.
Additional CO emission is found in the center of the galaxy and in the dust
cloud in the southern part of this image (Welch et al. 1998; Young 1998).
\label{205dust}
}
\end{figure}

The molecular clouds in NGC~185 and NGC~205 are very similar to 
Galactic giant molecular clouds.
Figures~\ref{185dust} and \ref{205dust} present images of CO emission, dust,
and HI in the dwarf ellipticals. 
The molecular clouds in the dwarf ellipticals have diameters about 60 pc, 
line widths ($\sigma$) of
3--8 km~s$^{-1}$, and therefore virial masses of $10^6 \rm M_\odot$ (Young
\& Lo 1996, 1997), which
are similar to Galactic clouds (e.g. Blitz 1993).
The molecular clouds in these figures are associated with HI clumps,
most probably photodissociated envelopes around the molecular gas, and with
dust clouds of extinction $A_V \approx\ 1$ mag.
Observations of the northern molecular cloud in NGC~205, made with the IRAM
30m telescope (Young 1998), have detected
$^{12}$CO~2-1, $^{12}$CO~1-0, and $^{13}$CO~1-0, and provide upper limits for
HCN~1-0 and HCO+~1-0.
The line ratios indicate optically thick, thermalized CO emission from
gas at moderate density, again very typical of Galactic molecular clouds.

An unexpected property of the ISM in NGC~185 and NGC~205 is the remarkably
low HI column densities--- only a few $\times$
$10^{20}$~cm$^{-2}$ and less.
When molecular clouds are observed in our own galaxy, they are associated
with HI column densities of about $10^{21}$ cm$^{-2}$ (Blitz 1993; Savage 1977).
Similar effects are observed in M31 (Lada et al. 1988).
The most common conclusion is that in our galaxy and M31, an 
HI column density of about $10^{21}$~cm$^{-2}$ is
necessary to shield molecular gas against the dissociating interstellar UV field.
The HI column densities in NGC~185 and NGC~205 would seem to be too low to
allow molecular gas to form.
However, they can be explained
by theoretical models of photodissociated regions (e.g. Draine \& Bertoldi
1996), if the ratio of the interstellar UV field (in units of the Habing
field) to the hydrogen volume density is around $10^{-2}$ or lower.
In comparison, a value of 0.05 is estimated for the Galactic star
formation region NGC~2023, which is illuminated by only one early B star
(Draine \& Bertoldi 1996).
We have no good density estimates in the dwarf ellipticals. 
However, Welch et al. (1996, 1998) estimate the interstellar UV fields in
NGC~185 and NGC~205, and they find that the fields are probably a factor of
10 to 100 lower than the solar neighborhood UV fields. 
The molecular clouds in NGC~185 and NGC~205 thus provide enlightening
comparisons with the ISM in our own spiral Galaxy.

\section{An optimistic outlook on star formation in early-type galaxies}

In Section~\ref{kinematics} we speculated that the neutral ISM in NGC~185
and NGC~205 may have an internal origin.
This speculation in turn suggests that a large fraction of these kinds of
early-type galaxies, especially in the field or in loose groups,
could have significant amounts of neutral gas.
Consideration of the molecular cloud properties in NGC~185 and NGC~205 also
suggests that the transformation of neutral gas into stars may be easier
than we thought, both in dwarf and in giant ellipticals.
The conditions within elliptical galaxies are not likely to be the same as
the conditions within spiral galaxies.
HI column densities which, by the standards of our Galaxy, are thought 
to be too low to support star formation, might not actually be too low.

\acknowledgments
Thanks to K. Y. Lo, G. Welch, J. van Gorkom, and M. Rupen for interesting
discussions.

\begin{question}{Paul Eskridge}
I'd like to clarify once more that NGC~185 and NGC~205 are not elliptical
galaxies.  They define the bright end of the L$_B$--concentration relation
that extends down to the Galactic halo dwarf spheroidals, not the relation
that is defined by bright ellipticals.
\end{question}

\begin{question}{Patricia Carral}
Can you tell us what is the gas-to-dust ratio in NGC~185 and NGC~205?
\end{question}
\begin{answer}{Young}
Table~\ref{table1} contains the best current estimates of dust and gas
masses in the dwarf ellipticals.
Here's another comparison: for the darker optical dust patch in NGC~185,
Price (1985) estimates a peak A$_V$ of 1.6 mag.
The total hydrogen column density at this point is a few $\times 10^{21}$
nuclei~cm$^{-2}$.
\end{answer}

\begin{question}{Paul Goudfrooij}
There are a number of bright young stars in the center of
NGC~205.  
Did you try to evaluate the UV radiation field from those stars to check
that it is indeed significantly lower than in a typical Galactic
star-forming region?
\end{question}
\begin{answer}{Young}
Yes, Welch's estimates (Welch et al. 1996, 1998) include both the young stars
and the old stars.  
The component from the old stars is scaled from observations of NGC~1399.
The young massive stars contribute some UV radiation, but they are not
tremendously massive.  I think they are B stars.
\end{answer}

\begin{question}{Gary Welch}
Those calculated UV radiation fields depend very much on unknown geometry
and on the 
assumed effective temperatures of the young stars, which are poorly known.
Also, if the radiation fields are low as we think they are, there
might also be a significant mass of diffuse molecular gas which is not
associated with known optical dust clouds. 
We have recently published NRAO 12m CO(1-0) observations which seem to show
such gas.
The total mass could exceed that inside the optical dust clouds, but
the detections need to be confirmed.
\end{question}

\end{document}